\documentclass[aps,prl,reprint,superscriptaddress]{revtex4-1}

\usepackage{hyperref}
\usepackage{xr-hyper}
\usepackage{graphicx}
\usepackage{etoolbox}
\usepackage{mathtools}

\renewcommand{\d}{\mathrm{d}}
\newcommand{\Vimp}{V_\text{imp}}
\newcommand{\reson}[2]{#1\!:\!#2}

\graphicspath{{images/}}

\newtoggle{includesupplementary}
\toggletrue{includesupplementary}

\begin{document}

\title{Strong quantum scarring by local impurities}

\author{Perttu J. J. Luukko}
\email{perttu.luukko@iki.fi}
\affiliation{Nanoscience Center, Department of Physics, University of Jyv\"askyl\"a, Jyv\"askyl\"a FI-40014, Finland}
\author{Byron Drury}
\affiliation{Department of Physics, Massachusetts Institute of Technology, Cambridge, Massachusetts 02139, USA}
\author{Anna Klales}
\affiliation{Department of Physics, Harvard University, Cambridge, Massachusetts 02138, USA}
\author{Lev Kaplan}
\affiliation{Department of Physics and Engineering Physics, Tulane University, New Orleans, Louisiana 70118, USA}
\author{Eric J. Heller}
\affiliation{Department of Physics, Harvard University, Cambridge, Massachusetts 02138, USA}
\author{Esa R\"as\"anen}
\affiliation{Department of Physics, Tampere University of Technology, Tampere FI-33101, Finland}

\begin{abstract}
We discover and characterize strong \emph{quantum scars}, or eigenstates resembling
classical periodic orbits, in two-dimensional quantum
wells perturbed by local impurities.
These scars are not explained by ordinary scar theory, which would require the
existence of short, moderately unstable periodic orbits in the perturbed
system.
Instead, they are supported by classical resonances in the unperturbed system
and the resulting quantum near-degeneracy.
Even in the case of a large number of randomly scattered impurities, the scars
prefer distinct orientations that extremize the overlap with the impurities.
We demonstrate that these preferred orientations can be used for highly efficient
transport of quantum wave packets across the perturbed potential landscape.
Assisted by the scars, wave-packet recurrences are significantly stronger than
in the \emph{unperturbed} system.
Together with the controllability of the preferred orientations, this property
may be very useful for quantum transport applications.
\end{abstract}

\date{\today}


\makeatletter
\hypersetup{pdfauthor={Perttu J. J. Luukko, Byron Drury, Anna Klales, Lev Kaplan, Eric J. Heller, Esa R\"as\"anen},pdftitle={\@title}}
\makeatother

\maketitle


Quantum scars \cite{scarsreview} are enhancements of probability density in the
eigenstates of a quantum chaotic system that occur around short unstable
periodic orbits (POs) of the corresponding classical system.
Scars have been observed experimentally in, e.g., microwave
cavities~\cite{sridhar,*stein}, optical cavities~\cite{sblee,*harayama}, and
quantum wells~\cite{fromhold,*wilkinson}, and computationally in, e.g., simulations of
graphene flakes~\cite{huang} and ultracold atomic gases~\cite{larson}.

Before the existence of scars was reported by Heller~\cite{scars}, eigenstates
of a classically chaotic system were conjectured to fill the available phase
space evenly, up to random fluctuations and energy conservation.
If high-energy eigenstates of non-regular (i.e., generic) systems were indeed
featureless and random, controlled applications in that regime would be
difficult.
Scars are therefore both a striking visual example of classical-quantum
correspondence away from the usual classical limit, and a useful example of a
quantum suppression of chaos.

In this work we describe quantum scars present in otherwise separable systems
disturbed by local perturbations such as impurity atoms.
In this case, scars are formed around POs of the corresponding
\emph{unperturbed} system.
These scars are significantly stronger than what ordinary scar theory predicts,
which we explain by introducing a new scarring mechanism.

In the following, values and equations are given in natural units where the
quantum Hamiltonian is simply $H = -\tfrac{1}{2}\nabla^2 + V + \Vimp$, where
$V$ is the unperturbed potential and $\Vimp$ represents the perturbation.

\vfill 

\section{Model system}
\label{sec:modelsystem}

The scarring mechanism, explained later in this Letter, is very general; it
requires only that the unperturbed system is separable, and that the perturbing
impurities are sufficiently local.
In the following we focus, for simplicity, on a few prototypical examples of a
circularly symmetric, two-dimensional potential well~$V(r)$ perturbed by
randomly scattered Gaussian bumps.

The classical POs of any circularly symmetric $V(r)$ can be enumerated directly
(see Supplementary Material and Ref.~\cite{orbits}).
Each PO is associated with a resonance, where the oscillation
frequencies of the radial and angular motion are commensurable.
The PO structure is especially simple if~$V(r)$ is a homogeneous function
(i.e., $V(r) \propto r^a$), since then POs with different total energies differ
only by a scaling.

In the following we use $a=5$.
Its shortest non-trivial \footnote{Besides the circular orbit and the case with
zero angular momentum} PO is a five-pointed star, which is both easily
distinguishable and short.
This PO corresponds to a $\reson{2}{5}$ classical resonance, where the orbit circles the
origin twice in the time of five radial oscillations.
The case of $a=2$, the harmonic oscillator, is special and ill-suited for
drawing general conclusions, while for $a=1, 3, 4$ the shortest non-trivial
POs are longer.

For easier comparison between the results, we focus in the following on a
single example shown in Fig.~\ref{fig:penta-examples}(a), where Gaussian bumps
with amplitude~$M=24$ are distributed in an unperturbed potential
$V(r)=\frac{1}{2}r^5$.
The average density of the bumps is two per unit square, and thus for a typical
energy of~$500$, approximately a hundred bumps exist in the classically
allowed region.
The full width at half maximum (FWHM) of the Gaussian bumps is $0.235$, which
is similar to the typical local wavelength of the eigenstates we consider.

The amplitude of the bumps is small compared to the total energy, making each
individual bump a small perturbation.
Nevertheless, together the impurities are sufficient to destroy classical
long-time stability; any stable structures present in the
otherwise chaotic Poincar\'e surface of section are tiny compared to $\hbar=1$.

\section{Scar observations}

\begin{figure}[htb]
  \centering
  \includegraphics[width=0.9\columnwidth]{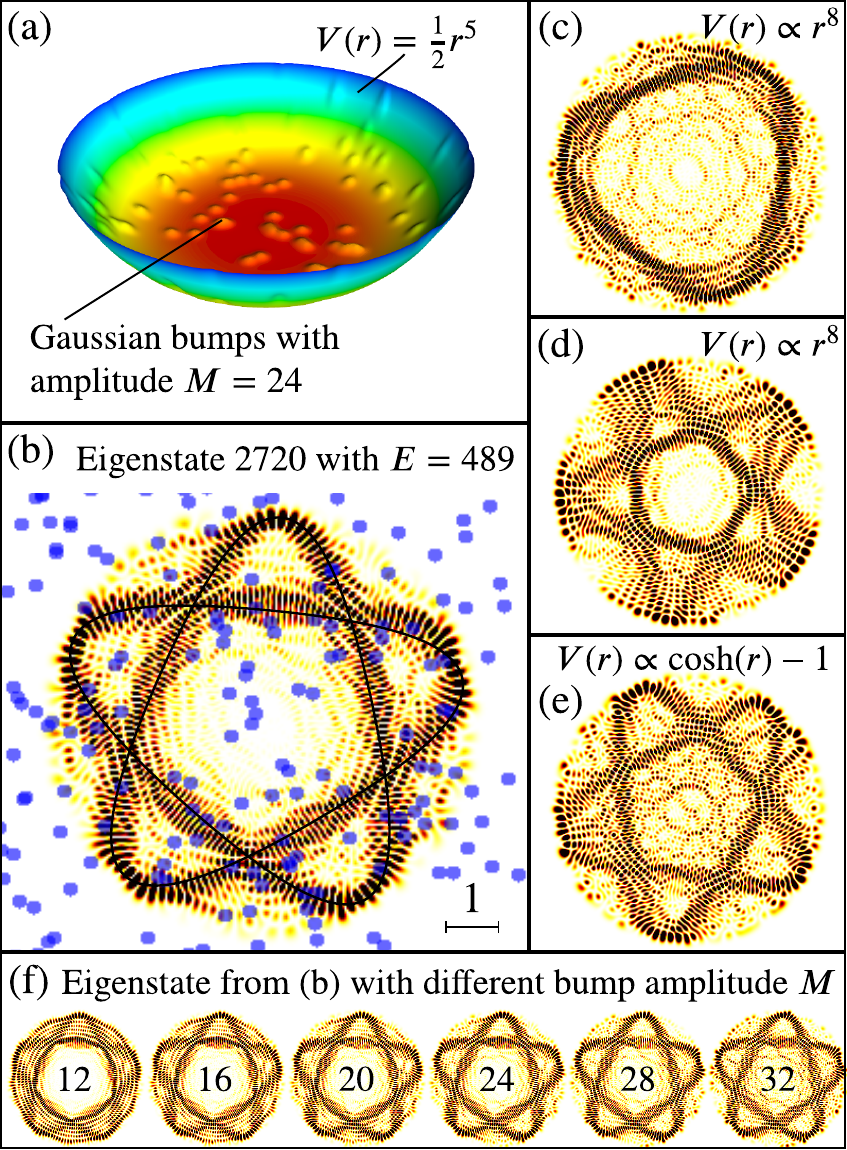}
  \caption{
  Examples of scarred eigenstates in perturbed potential wells. An example
  potential well $V(r) = \tfrac{1}{2}r^5$ perturbed by Gaussian bumps is shown
  in (a) and one of its strongly scarred eigenstates is shown in (b). Blue
  markers denote the locations and full widths at half maximum of the bumps,
  and the corresponding PO of the unperturbed potential is drawn as a solid
  line. Note that several bumps are located on the scar path. Examples of scars
  in other potentials are shown in (c)--(e). The eigenstate in (b) at different
  bump amplitudes~$M$ is shown in (f). The scar shape and orientation remain
  unchanged as $M$ increases.}
  \label{fig:penta-examples}
\end{figure}

Figure~\ref{fig:penta-examples}(b) shows an example of a scar found in the
eigenstates of the example potential described previously.
The eigenstates are solved with imaginary time propagation in real
space~\cite{itp2d}.
Figures~\ref{fig:penta-examples}(c)--(e) show examples of scars in a homogeneous
potential with exponent~$a=8$, and in a non-homogeneous potential $V(r) \propto
\cosh(r)-1$.
In all cases the scars follow POs in the corresponding unperturbed system.
We note that the scars are not a rare occurrence; for the potential in
Fig.~\ref{fig:penta-examples}(a) over 10\% of the eigenstates are clearly
scarred.
The scars are also very strong; in Fig~\ref{fig:penta-examples}(b) approximately
80\% of the probability density resides on the star path.

In ordinary scar theory \cite{scars}, each scar corresponds to
a moderately unstable PO in the classical system.
In this case such orbits do not exist.
For example, the shortest and least unstable PO near the scar shown in
Fig.~\ref{fig:penta-examples}(f) for $M=16$ closes on itself after two rounds
around the scar, and has a one-period stability exponent~\cite{gutz} $\chi
\approx 5$. This is by far too unstable to cause a conventional scar as strong.

Ordinary scar theory is also excluded by the behavior of the scars as a
function of the bump amplitude~$M$.
If $M$ is increased while keeping $\Vimp$ otherwise unchanged, the
scars grow stronger and then fade away without changing their orientation, as
shown in Fig.~\ref{fig:penta-examples}f.
A scar caused by ordinary scar theory should become rapidly weaker, since the
stability exponent of a PO should increase with~$M$.

Comparing scars at different energies~$E$ reveals that they occur in
only a few distinct orientations, and these orientations change quite slowly
with~$E$ (see Fig.~\ref{fig:fair-overlaps}).
For the example~$\Vimp$ used here there are three preferred
orientations. For other impurity realizations the number and location of the
preferred orientations vary (see Supplementary Material),
but the existence of preferred scar orientations is a generic feature.
This too contradicts predictions of ordinary scar theory.

\section{Wave-packet analysis}

Gaussian wave packets are a standard tool for studying eigenstate
scarring~\cite{scars}.
If a Gaussian wave packet $|\phi\rangle$ is centered on a scarring PO, its
autocorrelation function $A(t) = \langle\phi(0)|\phi(t)\rangle$ shows clear
short-time recurrences with a period~$T$ matching the period of the PO.
In ordinary scarring the strength of these recurrences $|A(nT)|^2$ dies out
as $1/\cosh(\chi n)$, where $\chi$ is the stability exponent of the
PO \cite{measuringscars}.
In addition, $|\langle\phi(0)|\psi\rangle|$ is large for an eigenstate scarred
by the particular PO, making wave packets useful for detecting scarred
eigenstates.

\begin{figure}[htb]
  \centering
  \includegraphics[width=\columnwidth]{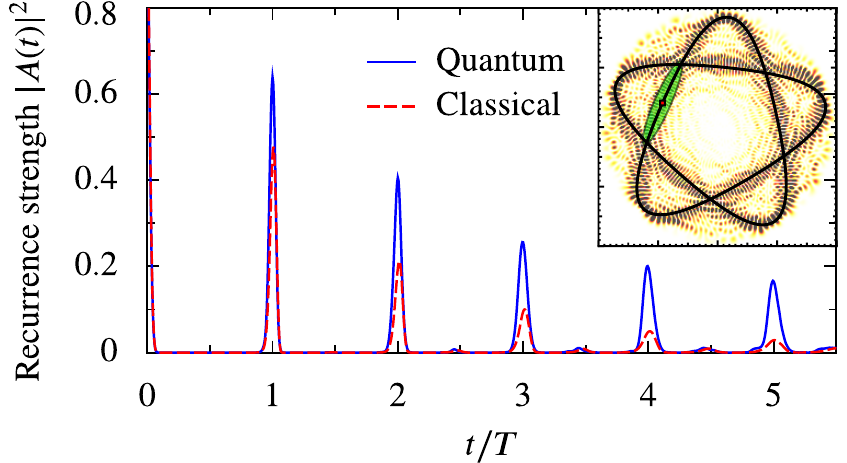}
  \caption{Recurrence strength of a Gaussian wave packet propagating along the
  scar in Fig.~\ref{fig:penta-examples}b. The quantum recurrence strength
  (solid blue line) is compared with the corresponding classical recurrence
  strength (dashed red line). Time is shown in units of the period~$T$ of the
  unperturbed PO. The scar, the corresponding PO (solid black line) and the
  location and full width at half maximum of the initial Gaussian (green area)
  are shown in the inset.}
  \label{fig:propagated-overlaps}
\end{figure}

In our case we use the PO of the unperturbed system to initialize the wave
packet.
The energy and the orientation of the PO are matched to the scarred eigenstate.
The width parameters of the Gaussian are matched approximately to the geometry
of the scar (see the inset in Fig.~\ref{fig:propagated-overlaps}).

The wave packets selected in this way have a considerable energy uncertainty,
so that many scars, sharing the same approximate orientation, contribute to the
recurrences.
A typical FWHM of the wave packet is approximately~$50$ energy units, or~$400$
eigenstates.

Figure~\ref{fig:propagated-overlaps} shows the recurrence strength $|A(t)|^2$
for a wave packet traveling on the scar shown in
Fig.~\ref{fig:penta-examples}(b).
Clear periodic recurrences are visible, with a period that matches the
period of the unperturbed PO.

To account for purely classical effects, we compare the results with the
recurrence of the corresponding classical density \footnote{This is given by the
Wigner transform~$G$ of the wave packet. Its recurrence strength was calculated
by sampling 80\,000 classical initial states $\{(\mathbf{r}_i,
\mathbf{p}_i)\}_i$ from the distribution~$G$, propagating each in time, and
computing $I(t) = \sum_i G(\mathbf{r}_i(t), \mathbf{p}_i(t))$. Classical time
integration was performed with the sixth-order symplectic integrator of
Blanes~\&~Moan~\cite{blanesmoan}. Once normalized so that $I(0) = 1$, $I(t)$
corresponds to the quantum recurrence strength.}.
The classical recurrences are significantly weaker than the quantum recurrences
even at $t=T$, and this difference grows rapidly at later times, illustrating
the quantum nature of the phenomenon.

\begin{figure}[htb]
  \centering
  \includegraphics[width=\columnwidth]{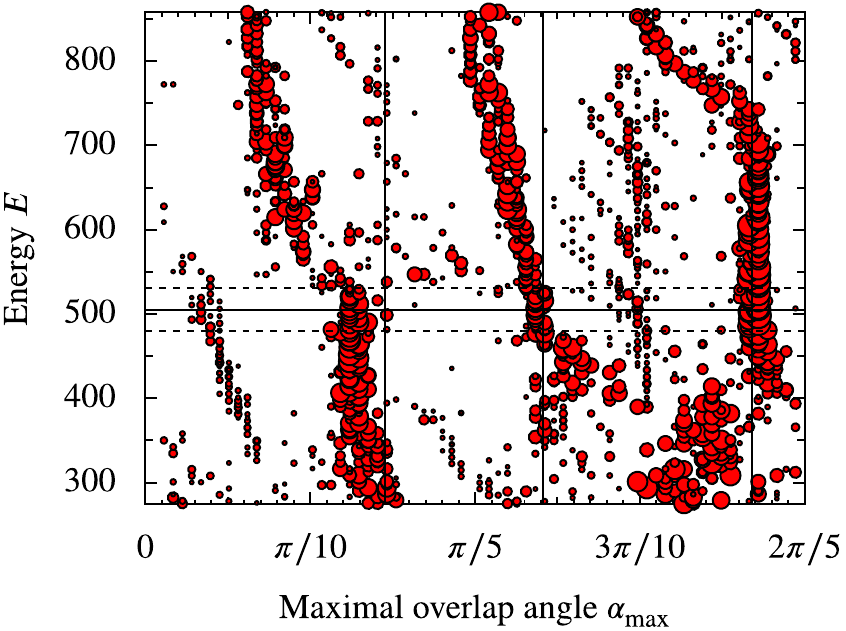}
  \caption{Scatter plot demonstrating the existence and stability of the
  preferred scar orientations. For each eigenstate, a circle is marked on the
  orientation angle~$\alpha$ showing the highest overlap between the eigenstate
  and an initial Gaussian wave packet with the given orientation. The radius of
  the circle gives its squared magnitude. Squared overlaps of less than
  $3\cdot10^{-3}$ are excluded. The vertical coordinate is the common energy of
  the eigenstate and the PO. The solid and dashed horizontal lines show,
  respectively, the mean energy and energy FWHM of the wave packet used in
  Fig.~\ref{fig:propagated-overlaps}. For easier comparison, orientation angles
  with peaks at $t=4T$ in Fig.~\ref{fig:angle-comparison}, are marked with
  solid vertical lines.}
  \label{fig:fair-overlaps}
\end{figure}

The existence of stable preferred scar orientations can be demonstrated by
systematically detecting scarred states with wave packets.
Figure~\ref{fig:fair-overlaps} shows how the overlap of the initial wave packet
with the target eigenstate depends on the energy of the eigenstate and the
orientation of the PO used to initialize the wave packet.
The orientation coordinate~$\alpha$ is such that at~$\alpha=0$ the
wave packet starts on the positive $y$-axis and heads to the right.

In an angular window of $2\pi/5$ (after which the PO is the same), three
branches of high overlaps are visible, corresponding to the preferred
orientations.
Note that the rightmost branch is roughly vertical for $E=400\dotsc700$,
corresponding to an increase of the average radius of the PO by roughly~$0.3$
units, which is the length scale of the individual bumps.

\begin{figure}[htb]
  \centering
  \includegraphics[width=\columnwidth]{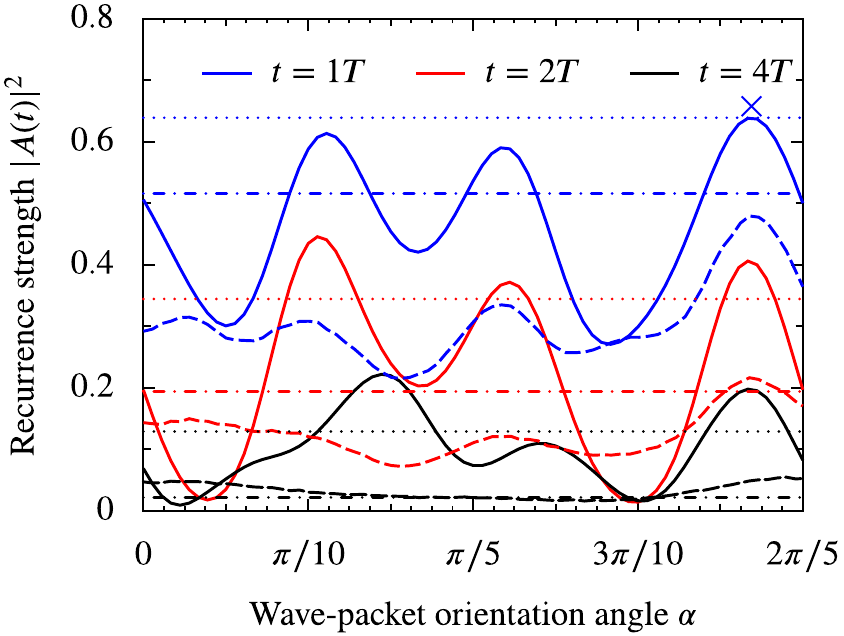}
  \caption{Amplitude of recurrence peaks shown in
  Fig.~\ref{fig:propagated-overlaps} as a function of the initial wave packet
  orientation~$\alpha$. Solid lines and dashed lines show the quantum and classical results,
  respectively. Dash-dotted and dotted lines show the quantum and classical
  results in the unperturbed system, respectively. Blue, red, and
  black lines correspond to snapshots after one,
  two, and four periods, respectively. The orientation angle used in
  Fig.~\ref{fig:propagated-overlaps} is marked with a blue cross. This is also
  the orientation that matches the scar shown in
  Fig.~\ref{fig:penta-examples}(b).}
  \label{fig:angle-comparison}
\end{figure}

Figure~\ref{fig:angle-comparison} shows how the amplitude of the recurrence
peaks in Fig.~\ref{fig:propagated-overlaps} depend of the orientation
angle~$\alpha$.
Both quantum and classical wave-packets show stronger short-time recurrences at
the preferred orientations, indicating that the preferred orientations can also
be explained classically.
However, especially at the preferred orientations, quantum late-time
recurrences are much stronger than in the classical case.
Note that the quantum recurrences are stronger than the classical ones even on
average, suggesting that there is also an effect that strengthens quantum
recurrences at all orientations.

To highlight how strong the quantum recurrences are, a comparison to
recurrences in the unperturbed system is also shown in
Fig.~\ref{fig:angle-comparison}.
For the preferred orientations, the quantum late-time recurrences greatly
exceed the strength of both the quantum and the classical recurrences in the
\emph{unperturbed} system.
Via the creation of strong scars with stable preferred orientations, the
randomly scattered impurities \emph{enhance} the coherent propagation of
quantum wave packets in the potential well!

\section{Perturbation theory}
\label{sec:theory}

Both the existence of scars and the preferred orientations can be explained by
perturbation theory.
This explanation is based on two ingredients. Firstly, special
nearly-degenerate subspaces exist in the basis of unperturbed eigenstates.
Secondly, the local perturbations select scarred eigenstates from these
subspaces.

The eigenstates of the unperturbed, circularly symmetric system are labeled by
two quantum numbers~$(r, m)$, corresponding to radial and angular motion,
respectively.
States $(r, \pm m)$ are exactly degenerate, but in addition there are
near-degeneracies that correspond to classical POs.

By the Bohr--Sommerfeld quantization condition (a good approximation at high
quantum numbers) the energy difference from increasing~$r$ or~$m$ by one is
proportional to the classical oscillation frequency of the corresponding
action.
Thus, if a state~$(r, m)$ is nearby in action to a classical PO with a ratio
$\reson{a}{b}$ between the oscillation frequencies, the state $(r+a, m-b$) will be
nearby in energy.
The smaller $a$ and $b$ are the closer the near-degeneracy is.
This creates ``resonant sets'' of unperturbed basis states, and a part
of a resonant set can be almost degenerate.
Expanding in the unperturbed basis reveals that the scarred eigenstates are
localized to such near-degenerate subspaces.

A superposition of two resonant states will exhibit beating in both the radial
and angular directions.
Because the ratio of the beat frequencies is also $\reson{a}{b}$, the interference
pattern will trace out the shape of the classical PO.
Adding more resonant states with appropriate phases will narrow the region of
constructive interference and sharpen the scar, and even a few basis states can
create a distinctly classical-looking linear combination.
Similar reconstruction of classical-like states from (nearly) degenerate basis
states has also been studied previously~\cite{liu, *squarelocalization,
*ripple, *pollet}.
However, to create strongly scarred eigenstates with preferred orientations a
mechanism that favors these scarred linear combinations is required.

Within the impurity strength regime that results in the strongest scarring, the
perturbation~$\Vimp$ mostly only couples resonant, near-degenerate basis
states.
This locality is enhanced by the rapid weakening of the coupling with
increasing difference in quantum numbers between the states.
One can therefore approximate the perturbed eigenstates by degenerate
perturbation theory (DPT), i.e., by diagonalizing $\Vimp$ within the
near-degenerate subspace.

The DPT-produced eigenstates corresponding to the extremal eigenvalues of
$\Vimp$ are, by the variational principle, the states $|\psi\rangle$ that
extremize the expectation value $\langle\Vimp\rangle \coloneqq
\langle\psi|\Vimp|\psi\rangle$.
This quantity for a non-scarred, spatially delocalized state is essentially an
average value of $\Vimp$ over the entire accessible region, and never an
extremum.
Compared to the non-scarred states, the scarred states in a near-degenerate
subspace have their probability amplitude concentrated in a much smaller region
of space.
For a $\Vimp$ consisting of local impurities the state that maximizes
(minimizes) $\langle\Vimp\rangle$ will therefore generally be
a scarred state oriented so as to coincide with an anomalously many (few)
impurities.

By the previous argument the scar orientations are mostly selected by the
positions of the impurities.
Since the inner and outer radii of the POs change slowly with energy, the
orientations that extremize $\langle\Vimp\rangle$ will be determined largely by
the same impurities for many different resonant sets.
This is the origin of the stability of the preferred orientations seen in
Fig.~\ref{fig:fair-overlaps}.

\begin{figure}[htb]
  \centering
  \includegraphics[width=\columnwidth]{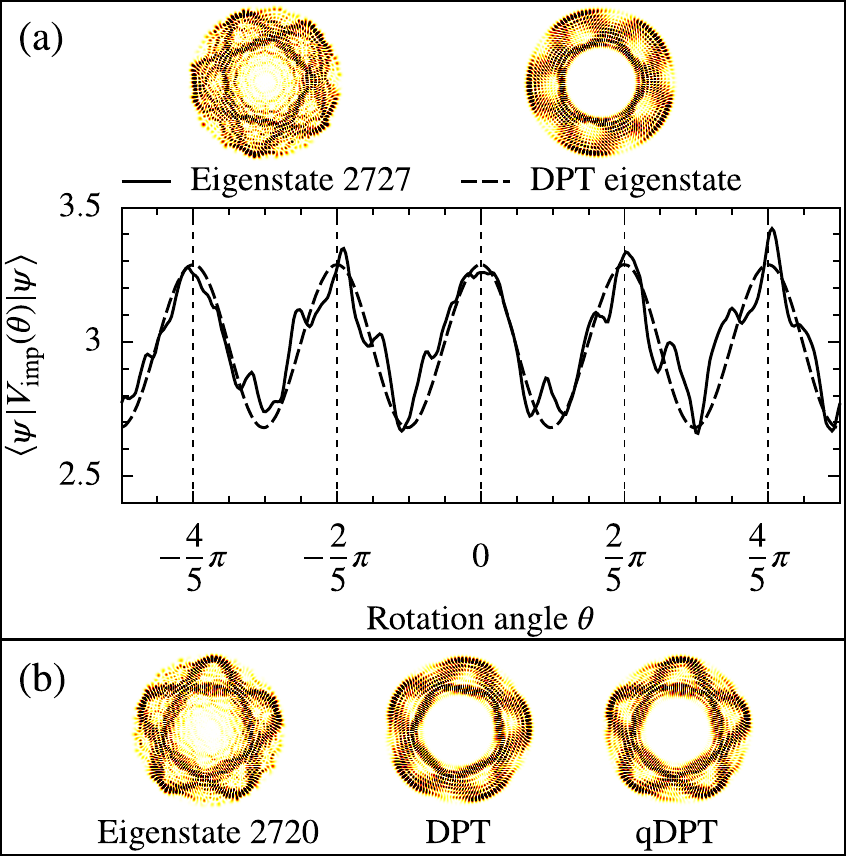}
  \caption{%
  Reconstructing scarred eigenstates with degenerate perturbation theory (DPT).
  DPT reconstruction of a scarred eigenstate $2727$ is shown in (a). The plot
  depicts the expectation value $\langle\Vimp(\theta)\rangle$, rotated by
  angle~$\theta$ from the original $\Vimp$, for both the actual eigenstate (solid line)
  and the eigenstate as predicted by DPT (dashed line). The plot shows that
  the orientation of the scar is such that $\langle\Vimp\rangle$ is maximized.
  The overlap between the true eigenstate and the DPT reconstruction is 56\%.
  The DPT reconstruction is calculated from a resonant set of only three nearly
  degenerate basis state doublets. Similar reconstruction of the eigenstate
  $2720$, shown also in Fig.~\ref{fig:penta-examples}b, is shown in (b). The
  DPT reconstruction produces a scar, but not at exactly the correct
  orientation. An improved quasi-DPT approximation, which takes into account
  the imperfect degeneracy, corrects it. The overlap between the eigenstate and
  the reconstruction is 60\% for DPT and 68\% for qDPT. The DPT reconstruction
  uses three nearly degenerate doublets, while the qDPT reconstruction uses
  five.
  }
  \label{fig:perturbation-theory}
\end{figure}

Though a simple DPT approximation does not reproduce the true eigenstates
exactly, it both explains the scarring and predicts the orientations of the
observed scars well, as illustrated in Fig.~\ref{fig:perturbation-theory}.
Taking into account the imperfect degeneracy amounts to diagonalizing the full
Hamiltonian instead of only the perturbation \cite{Davydov}.
This means that in the matrix that is diagonalized the diagonal elements are
shifted by the (small) spacings of the near-degenerate basis states.
This ``quasi-DPT'' (qDPT) approximation improves the agreement between the DPT
approximation and the exact eigenstates, as shown in
Fig.~\ref{fig:perturbation-theory}(b).

\section{Summary and outlook}

To summarize, we have shown that a new type of quantum scarring is found in
separable systems perturbed by local impurities.
The scars are very strong, and they tend to occur in discrete preferred
orientations.
This allows wave packets to propagate through the perturbed system with higher
fidelity than in the unperturbed system.

The theoretical basis of the scarring is very general, requiring only classical
resonances and local perturbations.
Therefore the scarring should have consequences well beyond the simple models
discussed here.

The implications of the enhanced wave packet recurrences for quantum transport
will be an important area for future work.
In an experiment, local perturbations similar to the ones used here could be
generated by a conducting nanotip~\cite{nanotip1,*nanotip2,*nanotip3},
selecting particular scar orientations and enhancing the local conductance in a
controlled way.


\begin{acknowledgments}
P.J.J.L. thanks the Finnish Cultural Foundation, the Magnus Ehrnrooth
Foundation, and the Emil Aaltonen Foundation for financial support. L.K.
acknowledges support from the U.S. NSF under Grant No.\ 1205788, and E.R. from the Academy of Finland. A.K. acknowledges that this work was supported by the STC Center for Integrated Quantum Materials, NSF Grant No.\ DMR-1231319. We are grateful
to CSC -- the Finnish IT Center for Science -- for providing computational
resources for numerical simulations. We wish to thank Prof.\ Li~Ge for useful
discussions.
\end{acknowledgments}

\bibliography{luukkoscars-paper}

\begin{thebibliography}{26}%
\makeatletter
\providecommand \@ifxundefined [1]{%
 \@ifx{#1\undefined}
}%
\providecommand \@ifnum [1]{%
 \ifnum #1\expandafter \@firstoftwo
 \else \expandafter \@secondoftwo
 \fi
}%
\providecommand \@ifx [1]{%
 \ifx #1\expandafter \@firstoftwo
 \else \expandafter \@secondoftwo
 \fi
}%
\providecommand \natexlab [1]{#1}%
\providecommand \enquote  [1]{``#1''}%
\providecommand \bibnamefont  [1]{#1}%
\providecommand \bibfnamefont [1]{#1}%
\providecommand \citenamefont [1]{#1}%
\providecommand \href@noop [0]{\@secondoftwo}%
\providecommand \href [0]{\begingroup \@sanitize@url \@href}%
\providecommand \@href[1]{\@@startlink{#1}\@@href}%
\providecommand \@@href[1]{\endgroup#1\@@endlink}%
\providecommand \@sanitize@url [0]{\catcode `\\12\catcode `\$12\catcode
  `\&12\catcode `\#12\catcode `\^12\catcode `\_12\catcode `\%12\relax}%
\providecommand \@@startlink[1]{}%
\providecommand \@@endlink[0]{}%
\providecommand \url  [0]{\begingroup\@sanitize@url \@url }%
\providecommand \@url [1]{\endgroup\@href {#1}{\urlprefix }}%
\providecommand \urlprefix  [0]{URL }%
\providecommand \Eprint [0]{\href }%
\providecommand \doibase [0]{http://dx.doi.org/}%
\providecommand \selectlanguage [0]{\@gobble}%
\providecommand \bibinfo  [0]{\@secondoftwo}%
\providecommand \bibfield  [0]{\@secondoftwo}%
\providecommand \translation [1]{[#1]}%
\providecommand \BibitemOpen [0]{}%
\providecommand \bibitemStop [0]{}%
\providecommand \bibitemNoStop [0]{.\EOS\space}%
\providecommand \EOS [0]{\spacefactor3000\relax}%
\providecommand \BibitemShut  [1]{\csname bibitem#1\endcsname}%
\let\auto@bib@innerbib\@empty
\bibitem [{\citenamefont {Kaplan}(1999)}]{scarsreview}%
  \BibitemOpen
  \bibfield  {author} {\bibinfo {author} {\bibfnamefont {L.}~\bibnamefont
  {Kaplan}},\ }\href@noop {} {\bibfield  {journal} {\bibinfo  {journal}
  {Nonlinearity}\ }\textbf {\bibinfo {volume} {12}},\ \bibinfo {pages} {R1}
  (\bibinfo {year} {1999})}\BibitemShut {NoStop}%
\bibitem [{\citenamefont {Sridhar}(1991)}]{sridhar}%
  \BibitemOpen
  \bibfield  {author} {\bibinfo {author} {\bibfnamefont {S.}~\bibnamefont
  {Sridhar}},\ }\href@noop {} {\bibfield  {journal} {\bibinfo  {journal}
  {Phys.~Rev.~Lett.}\ }\textbf {\bibinfo {volume} {67}},\ \bibinfo {pages}
  {785} (\bibinfo {year} {1991})}\BibitemShut {NoStop}%
\bibitem [{\citenamefont {Stein}\ and\ \citenamefont
  {St\"ockmann}(1992)}]{stein}%
  \BibitemOpen
  \bibfield  {author} {\bibinfo {author} {\bibfnamefont {J.}~\bibnamefont
  {Stein}}\ and\ \bibinfo {author} {\bibfnamefont {H.-J.}\ \bibnamefont
  {St\"ockmann}},\ }\href@noop {} {\bibfield  {journal} {\bibinfo  {journal}
  {Phys.~Rev.~Lett.}\ }\textbf {\bibinfo {volume} {68}},\ \bibinfo {pages}
  {2867} (\bibinfo {year} {1992})}\BibitemShut {NoStop}%
\bibitem [{\citenamefont {Lee}\ \emph {et~al.}(2002)\citenamefont {Lee},
  \citenamefont {Lee}, \citenamefont {Chang}, \citenamefont {Moon},
  \citenamefont {Kim},\ and\ \citenamefont {An}}]{sblee}%
  \BibitemOpen
  \bibfield  {author} {\bibinfo {author} {\bibfnamefont {S.-B.}\ \bibnamefont
  {Lee}}, \bibinfo {author} {\bibfnamefont {J.-H.}\ \bibnamefont {Lee}},
  \bibinfo {author} {\bibfnamefont {J.-S.}\ \bibnamefont {Chang}}, \bibinfo
  {author} {\bibfnamefont {H.-J.}\ \bibnamefont {Moon}}, \bibinfo {author}
  {\bibfnamefont {S.~W.}\ \bibnamefont {Kim}}, \ and\ \bibinfo {author}
  {\bibfnamefont {K.}~\bibnamefont {An}},\ }\href@noop {} {\bibfield  {journal}
  {\bibinfo  {journal} {Phys.~Rev.~Lett.}\ }\textbf {\bibinfo {volume} {88}},\
  \bibinfo {pages} {033903} (\bibinfo {year} {2002})}\BibitemShut {NoStop}%
\bibitem [{\citenamefont {Harayama}\ \emph {et~al.}(2003)\citenamefont
  {Harayama}, \citenamefont {Fukushima}, \citenamefont {Davis}, \citenamefont
  {Vaccaro}, \citenamefont {Miyasaka}, \citenamefont {Nishimura},\ and\
  \citenamefont {Aida}}]{harayama}%
  \BibitemOpen
  \bibfield  {author} {\bibinfo {author} {\bibfnamefont {T.}~\bibnamefont
  {Harayama}}, \bibinfo {author} {\bibfnamefont {T.}~\bibnamefont {Fukushima}},
  \bibinfo {author} {\bibfnamefont {P.}~\bibnamefont {Davis}}, \bibinfo
  {author} {\bibfnamefont {P.~O.}\ \bibnamefont {Vaccaro}}, \bibinfo {author}
  {\bibfnamefont {T.}~\bibnamefont {Miyasaka}}, \bibinfo {author}
  {\bibfnamefont {T.}~\bibnamefont {Nishimura}}, \ and\ \bibinfo {author}
  {\bibfnamefont {T.}~\bibnamefont {Aida}},\ }\href@noop {} {\bibfield
  {journal} {\bibinfo  {journal} {Phys.~Rev.~E}\ }\textbf {\bibinfo {volume}
  {67}},\ \bibinfo {pages} {015207} (\bibinfo {year} {2003})}\BibitemShut
  {NoStop}%
\bibitem [{\citenamefont {Fromhold}\ \emph {et~al.}(1995)\citenamefont
  {Fromhold}, \citenamefont {Wilkinson}, \citenamefont {Sheard}, \citenamefont
  {Eaves}, \citenamefont {Miao},\ and\ \citenamefont {Edwards}}]{fromhold}%
  \BibitemOpen
  \bibfield  {author} {\bibinfo {author} {\bibfnamefont {T.~M.}\ \bibnamefont
  {Fromhold}}, \bibinfo {author} {\bibfnamefont {P.~B.}\ \bibnamefont
  {Wilkinson}}, \bibinfo {author} {\bibfnamefont {F.~W.}\ \bibnamefont
  {Sheard}}, \bibinfo {author} {\bibfnamefont {L.}~\bibnamefont {Eaves}},
  \bibinfo {author} {\bibfnamefont {J.}~\bibnamefont {Miao}}, \ and\ \bibinfo
  {author} {\bibfnamefont {G.}~\bibnamefont {Edwards}},\ }\href@noop {}
  {\bibfield  {journal} {\bibinfo  {journal} {Phys.~Rev.~Lett.}\ }\textbf
  {\bibinfo {volume} {75}},\ \bibinfo {pages} {1142} (\bibinfo {year}
  {1995})}\BibitemShut {NoStop}%
\bibitem [{\citenamefont {{Wilkinson}}\ \emph {et~al.}(1996)\citenamefont
  {{Wilkinson}}, \citenamefont {{Fromhold}}, \citenamefont {{Eaves}},
  \citenamefont {{Sheard}}, \citenamefont {{Miura}},\ and\ \citenamefont
  {{Takamasu}}}]{wilkinson}%
  \BibitemOpen
  \bibfield  {author} {\bibinfo {author} {\bibfnamefont {P.~B.}\ \bibnamefont
  {{Wilkinson}}}, \bibinfo {author} {\bibfnamefont {T.~M.}\ \bibnamefont
  {{Fromhold}}}, \bibinfo {author} {\bibfnamefont {L.}~\bibnamefont {{Eaves}}},
  \bibinfo {author} {\bibfnamefont {F.~W.}\ \bibnamefont {{Sheard}}}, \bibinfo
  {author} {\bibfnamefont {N.}~\bibnamefont {{Miura}}}, \ and\ \bibinfo
  {author} {\bibfnamefont {T.}~\bibnamefont {{Takamasu}}},\ }\href@noop {}
  {\bibfield  {journal} {\bibinfo  {journal} {Nature}\ }\textbf {\bibinfo
  {volume} {380}},\ \bibinfo {pages} {608} (\bibinfo {year}
  {1996})}\BibitemShut {NoStop}%
\bibitem [{\citenamefont {Huang}\ \emph {et~al.}(2009)\citenamefont {Huang},
  \citenamefont {Lai}, \citenamefont {Ferry}, \citenamefont {Goodnick},\ and\
  \citenamefont {Akis}}]{huang}%
  \BibitemOpen
  \bibfield  {author} {\bibinfo {author} {\bibfnamefont {L.}~\bibnamefont
  {Huang}}, \bibinfo {author} {\bibfnamefont {Y.-C.}\ \bibnamefont {Lai}},
  \bibinfo {author} {\bibfnamefont {D.~K.}\ \bibnamefont {Ferry}}, \bibinfo
  {author} {\bibfnamefont {S.~M.}\ \bibnamefont {Goodnick}}, \ and\ \bibinfo
  {author} {\bibfnamefont {R.}~\bibnamefont {Akis}},\ }\href@noop {} {\bibfield
   {journal} {\bibinfo  {journal} {Phys.~Rev.~Lett.}\ }\textbf {\bibinfo
  {volume} {103}},\ \bibinfo {pages} {054101} (\bibinfo {year}
  {2009})}\BibitemShut {NoStop}%
\bibitem [{\citenamefont {Larson}\ \emph {et~al.}(2013)\citenamefont {Larson},
  \citenamefont {Anderson},\ and\ \citenamefont {Altland}}]{larson}%
  \BibitemOpen
  \bibfield  {author} {\bibinfo {author} {\bibfnamefont {J.}~\bibnamefont
  {Larson}}, \bibinfo {author} {\bibfnamefont {B.~M.}\ \bibnamefont
  {Anderson}}, \ and\ \bibinfo {author} {\bibfnamefont {A.}~\bibnamefont
  {Altland}},\ }\href@noop {} {\bibfield  {journal} {\bibinfo  {journal}
  {Phys.~Rev.~A}\ }\textbf {\bibinfo {volume} {87}},\ \bibinfo {pages} {013624}
  (\bibinfo {year} {2013})}\BibitemShut {NoStop}%
\bibitem [{\citenamefont {Heller}(1984)}]{scars}%
  \BibitemOpen
  \bibfield  {author} {\bibinfo {author} {\bibfnamefont {E.~J.}\ \bibnamefont
  {Heller}},\ }\href {\doibase 10.1103/PhysRevLett.53.1515} {\bibfield
  {journal} {\bibinfo  {journal} {Phys.~Rev.~Lett.}\ }\textbf {\bibinfo
  {volume} {53}},\ \bibinfo {pages} {1515} (\bibinfo {year}
  {1984})}\BibitemShut {NoStop}%
\bibitem [{\citenamefont {Reynolds}\ and\ \citenamefont
  {Shouppe}(2010)}]{orbits}%
  \BibitemOpen
  \bibfield  {author} {\bibinfo {author} {\bibfnamefont {M.~A.}\ \bibnamefont
  {Reynolds}}\ and\ \bibinfo {author} {\bibfnamefont {M.~T.}\ \bibnamefont
  {Shouppe}},\ }\href@noop {} {\bibfield  {journal} {\bibinfo  {journal} {ArXiv
  e-prints}\ }\textbf {\bibinfo {volume} {1008.0559}} (\bibinfo {year}
  {2010})},\ \Eprint {http://arxiv.org/abs/1008.0559} {arXiv:1008.0559
  [physics.class-ph]} \BibitemShut {NoStop}%
\bibitem [{Note1()}]{Note1}%
  \BibitemOpen
  \bibinfo {note} {Besides the circular orbit and the case with zero angular
  momentum}\BibitemShut {NoStop}%
\bibitem [{\citenamefont {Luukko}\ and\ \citenamefont
  {R{\"a}s{\"a}nen}(2013)}]{itp2d}%
  \BibitemOpen
  \bibfield  {author} {\bibinfo {author} {\bibfnamefont {P.~J.~J.}\
  \bibnamefont {Luukko}}\ and\ \bibinfo {author} {\bibfnamefont
  {E.}~\bibnamefont {R{\"a}s{\"a}nen}},\ }\href@noop {} {\bibfield  {journal}
  {\bibinfo  {journal} {Comput.~Phys.~Commun.}\ }\textbf {\bibinfo {volume}
  {184}},\ \bibinfo {pages} {769} (\bibinfo {year} {2013})}\BibitemShut
  {NoStop}%
\bibitem [{\citenamefont {Gutzwiller}(1990)}]{gutz}%
  \BibitemOpen
  \bibfield  {author} {\bibinfo {author} {\bibfnamefont {M.~C.}\ \bibnamefont
  {Gutzwiller}},\ }\href@noop {} {\emph {\bibinfo {title} {Chaos in Classical
  and Quantum Mechanics}}}\ (\bibinfo  {publisher} {Springer},\ \bibinfo {year}
  {1990})\BibitemShut {NoStop}%
\bibitem [{\citenamefont {Kaplan}\ and\ \citenamefont
  {Heller}(1999)}]{measuringscars}%
  \BibitemOpen
  \bibfield  {author} {\bibinfo {author} {\bibfnamefont {L.}~\bibnamefont
  {Kaplan}}\ and\ \bibinfo {author} {\bibfnamefont {E.~J.}\ \bibnamefont
  {Heller}},\ }\href@noop {} {\bibfield  {journal} {\bibinfo  {journal} {Phys.
  Rev. E}\ }\textbf {\bibinfo {volume} {59}},\ \bibinfo {pages} {6609}
  (\bibinfo {year} {1999})}\BibitemShut {NoStop}%
\bibitem [{Note2()}]{Note2}%
  \BibitemOpen
  \bibinfo {note} {This is given by the Wigner transform~$G$ of the wave
  packet. Its recurrence strength was calculated by sampling 80\protect
  \tmspace +\thinmuskip {.1667em}000 classical initial states $\protect
  \{(\protect \mathbf {r}_i, \protect \mathbf {p}_i)\protect \}_i$ from the
  distribution~$G$, propagating each in time, and computing $I(t) = \DOTSB
  \sum@ \slimits@ _i G(\protect \mathbf {r}_i(t), \protect \mathbf {p}_i(t))$.
  Classical time integration was performed with the sixth-order symplectic
  integrator of Blanes~\&~Moan~\cite {blanesmoan}. Once normalized so that
  $I(0) = 1$, $I(t)$ corresponds to the quantum recurrence
  strength.}\BibitemShut {Stop}%
\bibitem [{\citenamefont {Liu}\ \emph {et~al.}(2006)\citenamefont {Liu},
  \citenamefont {Lu}, \citenamefont {Chen},\ and\ \citenamefont {Huang}}]{liu}%
  \BibitemOpen
  \bibfield  {author} {\bibinfo {author} {\bibfnamefont {C.~C.}\ \bibnamefont
  {Liu}}, \bibinfo {author} {\bibfnamefont {T.~H.}\ \bibnamefont {Lu}},
  \bibinfo {author} {\bibfnamefont {Y.~F.}\ \bibnamefont {Chen}}, \ and\
  \bibinfo {author} {\bibfnamefont {K.~F.}\ \bibnamefont {Huang}},\ }\href
  {\doibase 10.1103/PhysRevE.74.046214} {\bibfield  {journal} {\bibinfo
  {journal} {Phys.~Rev.~E}\ }\textbf {\bibinfo {volume} {74}},\ \bibinfo
  {pages} {046214} (\bibinfo {year} {2006})}\BibitemShut {NoStop}%
\bibitem [{\citenamefont {Chen}\ \emph {et~al.}(2002)\citenamefont {Chen},
  \citenamefont {Huang},\ and\ \citenamefont {Lan}}]{squarelocalization}%
  \BibitemOpen
  \bibfield  {author} {\bibinfo {author} {\bibfnamefont {Y.~F.}\ \bibnamefont
  {Chen}}, \bibinfo {author} {\bibfnamefont {K.~F.}\ \bibnamefont {Huang}}, \
  and\ \bibinfo {author} {\bibfnamefont {Y.~P.}\ \bibnamefont {Lan}},\ }\href
  {\doibase 10.1103/PhysRevE.66.046215} {\bibfield  {journal} {\bibinfo
  {journal} {Phys.~Rev.~E}\ }\textbf {\bibinfo {volume} {66}},\ \bibinfo
  {pages} {046215} (\bibinfo {year} {2002})}\BibitemShut {NoStop}%
\bibitem [{\citenamefont {Li}\ \emph {et~al.}(2002)\citenamefont {Li},
  \citenamefont {Reichl},\ and\ \citenamefont {Wu}}]{ripple}%
  \BibitemOpen
  \bibfield  {author} {\bibinfo {author} {\bibfnamefont {W.}~\bibnamefont
  {Li}}, \bibinfo {author} {\bibfnamefont {L.~E.}\ \bibnamefont {Reichl}}, \
  and\ \bibinfo {author} {\bibfnamefont {B.}~\bibnamefont {Wu}},\ }\href
  {\doibase 10.1103/PhysRevE.65.056220} {\bibfield  {journal} {\bibinfo
  {journal} {Phys.~Rev.~E}\ }\textbf {\bibinfo {volume} {65}},\ \bibinfo
  {pages} {056220} (\bibinfo {year} {2002})}\BibitemShut {NoStop}%
\bibitem [{\citenamefont {Pollet}\ \emph {et~al.}(1995)\citenamefont {Pollet},
  \citenamefont {M{\'e}plan},\ and\ \citenamefont {Gignoux}}]{pollet}%
  \BibitemOpen
  \bibfield  {author} {\bibinfo {author} {\bibfnamefont {J.}~\bibnamefont
  {Pollet}}, \bibinfo {author} {\bibfnamefont {O.}~\bibnamefont {M{\'e}plan}},
  \ and\ \bibinfo {author} {\bibfnamefont {C.}~\bibnamefont {Gignoux}},\
  }\href@noop {} {\bibfield  {journal} {\bibinfo  {journal} {J.~Phys.~A}\
  }\textbf {\bibinfo {volume} {28}},\ \bibinfo {pages} {7287} (\bibinfo {year}
  {1995})}\BibitemShut {NoStop}%
\bibitem [{\citenamefont {Davydov}(1976)}]{Davydov}%
  \BibitemOpen
  \bibfield  {author} {\bibinfo {author} {\bibfnamefont {A.~S.}\ \bibnamefont
  {Davydov}},\ }\href@noop {} {\emph {\bibinfo {title} {Quantum Mechanics}}},\
  \bibinfo {edition} {2nd}\ ed.\ (\bibinfo  {publisher} {Pergamon Press},\
  \bibinfo {year} {1976})\BibitemShut {NoStop}%
\bibitem [{\citenamefont {Bleszynski}\ \emph {et~al.}(2007)\citenamefont
  {Bleszynski}, \citenamefont {Zwanenburg}, \citenamefont {Westervelt},
  \citenamefont {Roest}, \citenamefont {Bakkers}, ,\ and\ \citenamefont
  {Kouwenhoven}}]{nanotip1}%
  \BibitemOpen
  \bibfield  {author} {\bibinfo {author} {\bibfnamefont {A.~C.}\ \bibnamefont
  {Bleszynski}}, \bibinfo {author} {\bibfnamefont {F.~A.}\ \bibnamefont
  {Zwanenburg}}, \bibinfo {author} {\bibfnamefont {R.~M.}\ \bibnamefont
  {Westervelt}}, \bibinfo {author} {\bibfnamefont {A.~L.}\ \bibnamefont
  {Roest}}, \bibinfo {author} {\bibfnamefont {E.~P. A.~M.}\ \bibnamefont
  {Bakkers}}, , \ and\ \bibinfo {author} {\bibfnamefont {L.~P.}\ \bibnamefont
  {Kouwenhoven}},\ }\href@noop {} {\bibfield  {journal} {\bibinfo  {journal}
  {Nano~Lett.}\ }\textbf {\bibinfo {volume} {7}},\ \bibinfo {pages} {2559}
  (\bibinfo {year} {2007})}\BibitemShut {NoStop}%
\bibitem [{\citenamefont {Boyd}\ \emph {et~al.}(2011)\citenamefont {Boyd},
  \citenamefont {Storm}, \citenamefont {Samuelson},\ and\ \citenamefont
  {Westervelt}}]{nanotip2}%
  \BibitemOpen
  \bibfield  {author} {\bibinfo {author} {\bibfnamefont {E.~E.}\ \bibnamefont
  {Boyd}}, \bibinfo {author} {\bibfnamefont {K.}~\bibnamefont {Storm}},
  \bibinfo {author} {\bibfnamefont {L.}~\bibnamefont {Samuelson}}, \ and\
  \bibinfo {author} {\bibfnamefont {R.~M.}\ \bibnamefont {Westervelt}},\
  }\href@noop {} {\bibfield  {journal} {\bibinfo  {journal} {Nanotechnology}\
  }\textbf {\bibinfo {volume} {22}},\ \bibinfo {pages} {185201} (\bibinfo
  {year} {2011})}\BibitemShut {NoStop}%
\bibitem [{\citenamefont {Blasi}\ \emph {et~al.}(2013)\citenamefont {Blasi},
  \citenamefont {Borunda}, \citenamefont {R\"as\"anen},\ and\ \citenamefont
  {Heller}}]{nanotip3}%
  \BibitemOpen
  \bibfield  {author} {\bibinfo {author} {\bibfnamefont {T.}~\bibnamefont
  {Blasi}}, \bibinfo {author} {\bibfnamefont {M.~F.}\ \bibnamefont {Borunda}},
  \bibinfo {author} {\bibfnamefont {E.}~\bibnamefont {R\"as\"anen}}, \ and\
  \bibinfo {author} {\bibfnamefont {E.~J.}\ \bibnamefont {Heller}},\
  }\href@noop {} {\bibfield  {journal} {\bibinfo  {journal} {Phys.~Rev.~B}\
  }\textbf {\bibinfo {volume} {87}},\ \bibinfo {pages} {241303} (\bibinfo
  {year} {2013})}\BibitemShut {NoStop}%
\bibitem [{\citenamefont {Blanes}\ and\ \citenamefont
  {Moan}(2002)}]{blanesmoan}%
  \BibitemOpen
  \bibfield  {author} {\bibinfo {author} {\bibfnamefont {S.}~\bibnamefont
  {Blanes}}\ and\ \bibinfo {author} {\bibfnamefont {P.~C.}\ \bibnamefont
  {Moan}},\ }\href@noop {} {\bibfield  {journal} {\bibinfo  {journal}
  {J.~Comput.~Appl.~Math.}\ }\textbf {\bibinfo {volume} {142}},\ \bibinfo
  {pages} {313} (\bibinfo {year} {2002})}\BibitemShut {NoStop}%
\bibitem [{\citenamefont {Goldstein}(1980)}]{goldstein}%
  \BibitemOpen
  \bibfield  {author} {\bibinfo {author} {\bibfnamefont {H.}~\bibnamefont
  {Goldstein}},\ }\href@noop {} {\emph {\bibinfo {title} {Classical
  Mechanics}}},\ \bibinfo {edition} {2nd}\ ed.\ (\bibinfo  {publisher}
  {Addison-Wesley},\ \bibinfo {year} {1980})\BibitemShut {NoStop}%
\end{thebibliography}%

%

\iftoggle{includesupplementary}{%
\appendix
\clearpage
\begin{center}
  \makeatletter
  \large\textbf{Supplementary material for ``\@title''}
  \makeatother
\end{center}

\section{Solving the periodic orbits of a circularly symmetric potential in 2D}
\label{supp:orbits}

The solution of the equations of motion for a classical particle in a
circularly symmetric potential~$V(r)$ can be found in standard texts on
classical mechanics (e.g., Chapter~3 in Ref.~\cite{goldstein}), but we summarize it here.
The total energy~$E$ of the particle is the sum of kinetic and potential
energy, which in polar coordinates reads (in units where mass $m=1$)
\begin{equation}
\label{eq:totalE}
E = \tfrac{1}{2}\dot{r}^2 + \frac{L^2}{2r^2} + V(r),
\end{equation}
where~$L$ is the angular momentum, which is also a constant of motion.
Solving for $\dot{r}$ from Eq.~\eqref{eq:totalE} gives
\begin{equation}
\label{eq:dotr}
\dot{r} = \pm\sqrt{2}\sqrt{E-V(r)-\frac{L^2}{2r^2}}.
\end{equation}
Changing variables in Eq.~\eqref{eq:dotr} from time to polar angle~$\phi$
(using $\dot\phi = L/r^2$) and inverting gives
\begin{equation}
\label{eq:phidiff}
\frac{\d\phi}{\d r} = \pm\frac{L}{r\sqrt{2r^2\left[E-V(r)\right]-L^2}} \coloneqq \pm\frac{L}{r\sqrt{f(r)}},
\end{equation}
where the function inside the square root is denoted as~$f$.
Equation~\eqref{eq:phidiff} can be conveniently integrated to give the
polar angle~$\phi$ as a function of the radial coordinate~$r$.

Assuming that~$V(r)$ is a potential well ($V(r)$ is monotonically
increasing and larger than~$E$ for large enough~$r$) and that $L \neq 0$, the radius of the particle oscillates between
two turning points $r=a$ and $r=b$, which can be solved from Eq.~\eqref{eq:dotr} by
setting $\dot{r} = 0$. Not surprisingly, these turning points are exactly where
$\d\phi/\d r$ diverges, i.e., the zeros of~$f$ in Eq.~\eqref{eq:phidiff}. Between
these turning points the radial coordinate either monotonically increases or
decreases, corresponding to the positive and negative solutions
in Eq.~\eqref{eq:dotr}.

When the particle completes one oscillation from $r=a$ to $r=b$ and back, the
corresponding change in the polar angle (by integrating Eq.~\eqref{eq:phidiff}) is
\begin{equation}
\label{eq:deltatheta}
\Delta\phi = \int_a^b \frac{2L}{r\sqrt{f(r)}}.
\end{equation}
Note that $\Delta\phi$ is completely specified by the form of the
potential~$V(r)$ and the values for~$E$ and~$L$. The particle eventually
returns to its starting point, i.e., the orbit is periodic, exactly when
$\Delta\phi$ is a rational multiple of~$2\pi$. If $\Delta\phi =
2\pi\frac{m}{n}$ for integer~$n$ and~$m$, after~$n$ oscillations between the
radial turning points, the particle has rotated around the origin~$m$ times,
returning exactly to its starting position with its original velocity, i.e., the classical oscillation frequencies are in $\reson{m}{n}$ resonance.
If the orbit is not periodic, it is quasiperiodic.

Searching for periodic orbits (POs) for a given potential~$V(r)$ and total energy~$E$
can be done conveniently by looking for zeros of the function
\(
g(E, L, n) \coloneqq \sin\left(\tfrac{1}{2}n\Delta\phi\right).
\)
This can be done by any standard root-finding method, using~$L$ as the free
variable, and some set of discrete choices for the integer~$n$. This gives
values of~$L$ which correspond to a PO, and the integers~$n$
and~$m$, which give the shape of the orbit.

The procedure is especially simple if~$V(r)$ is a homogeneous function of~$r$.
In this case
different total energies differ only by a scaling factor in space and time,
which means the PO shapes do not depend on~$E$. Moreover, for $V(r)
\propto r^a$ with integer~$a$, the function $f$ in Eq.~\eqref{eq:phidiff} is a
polynomial, which makes it particularly easy to find the turning points for moderate values of~$a$.

As an
illustration, the shortest POs for some small integer values of~$a$
are given in Table~\ref{tab:raorbits}. Most importantly, the five-pointed star
with $m/n = 2/5$ appears at~$a=5$, and the next jump to a simpler shortest
PO occurs at~$a=8$ with the birth of the triangle orbit with $m/n =
1/3$. The POs of power-law potentials are also studied in a more
indirect way in Ref.~\cite{orbits}.

\begin{table*}[htpb]
  \centering
  \caption{Periodic orbits for $V(r) \propto r^a$ up to $n=15$. Table
  entries are aligned by~$n$ to highlight how new PO shapes appear
  with increasing~$a$.}
  \begin{ruledtabular}
  \begin{tabular}{rrrrrrrrrrr}
	$a$ & \multicolumn{9}{c}{Periodic orbits as values of $m/n$}\\
	\hline
	1 &     &     & 4/7 &     & 5/9 & 6/11 &      & 7/13 & & 8/15\\
	2 & \multicolumn{9}{l}{1/2 (harmonic oscillator -- a very special case)}\\
	3 &     &     &     &     &     & 5/11 &      & 6/13 & & 7/15\\
	4 &     &     & 3/7 &     & 4/9 & 5/11 & 5/12 & 6/13 & & 7/15\\
	5 &     & 2/5 & 3/7 &     & 4/9 & 5/11 & 5/12 & \{5, 6\}/13 & & 7/15\\
	6 &     & 2/5 & 3/7 & 3/8 & 4/9 & \{4, 5\}/11 & 5/12 & \{5, 6\}/13 & 5/14 & 7/15\\
	7 &     & 2/5 & 3/7 & 3/8 & 4/9 & \{4, 5\}/11 & 5/12 & \{5, 6\}/13 & 5/14 & 7/15\\
	8 & 1/3 & 2/5 & 3/7 & 3/8 & 4/9 & \{4, 5\}/11 & 5/12 & \{5, 6\}/13 & 5/14 & 7/15\\
	9 & 1/3 & 2/5 & 3/7 & 3/8 & 4/9 & \{4, 5\}/11 & 5/12 & \{4, 5, 6\}/13 & 5/14 & 7/15\\
  \end{tabular}
  \end{ruledtabular}
  \label{tab:raorbits}
\end{table*}

\section{Extracting scars with a wave-packet ``scarmometer''}
\label{supp:scarmometer}

As explained in the main article, for ordinary scars a wave packet initialized
on a PO can be used to locate eigenstates that are scarred by that
particular orbit, since the scarred eigenstates will have large overlaps with
the wave packet. Similarly a wave packet initialized on a specific PO
of the unperturbed system can be used to isolate scars with that
particular orientation in the perturbed system. This is illustrated in
Fig.~\ref{fig:wavepacket-decomposition}, which uses the wave packet used in
Fig.~\ref{fig:propagated-overlaps} to isolate scarred
eigenstates with the same orientation as the $n = 2720$ example scar used
throughout the main article. This illustration also shows how several scarred
eigenstates contribute to the recurrences of a single wave packet.

\begin{figure*}[htbp]
  \centering
  \includegraphics[width=0.75\textwidth]{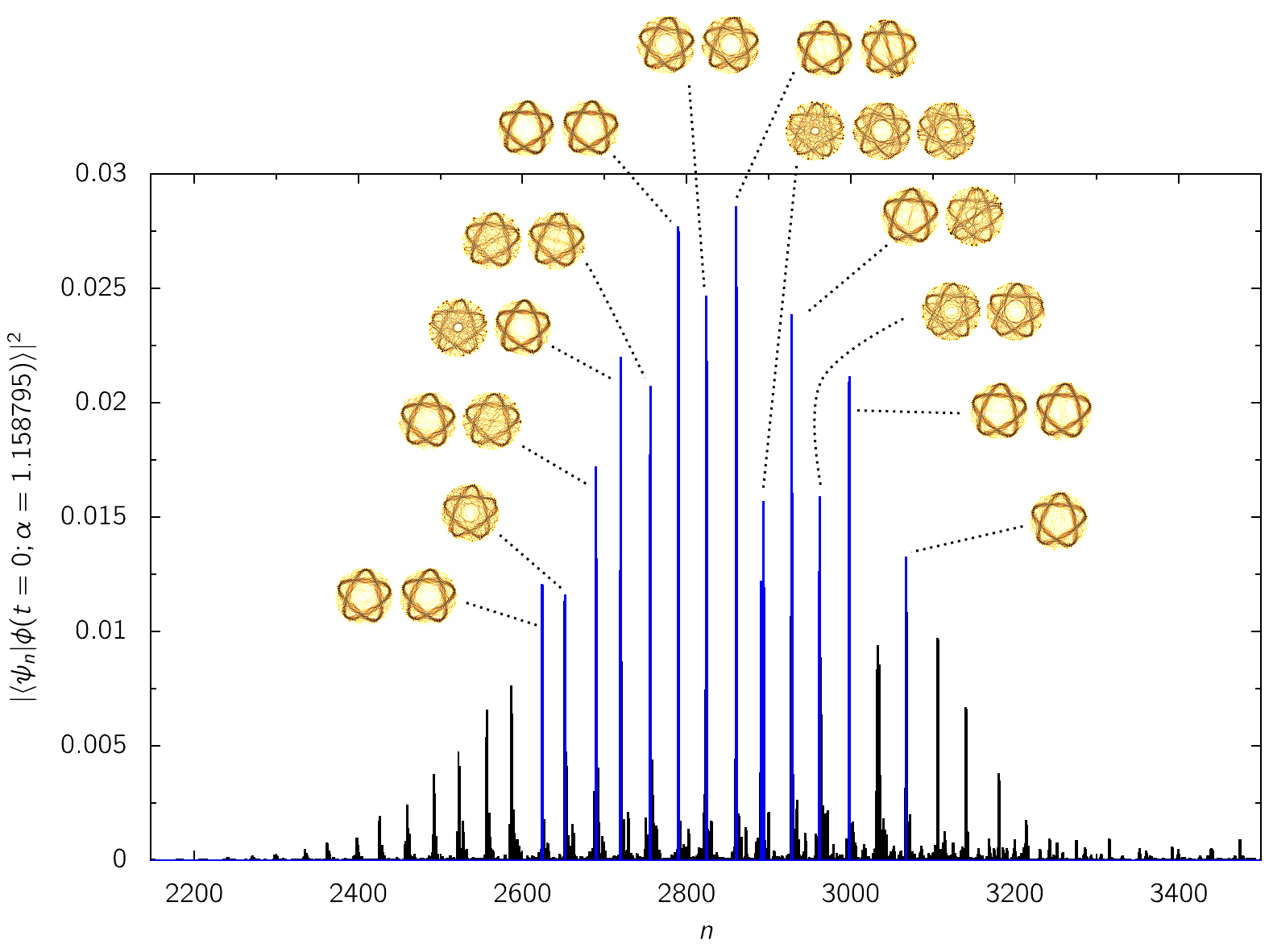}
  \caption{Extracting scars with a specific orientation with a wave packet
  ``scarmometer''. The histogram in the background shows the basis
  decomposition of the wave packet used in Fig.~\ref{fig:propagated-overlaps}
  in the eigenstates of the perturbed system. Scarred eigenstates show up as
  prominent peaks in this decomposition. For some of the strongest peaks
  (marked in blue) the corresponding eigenstate is also shown, highlighting
  that they are all scarred to a varying degree, and that the orientation of
  the five-pointed scar is the same.}
  \label{fig:wavepacket-decomposition}
\end{figure*}

\section{Scars of other resonances in homogeneous potentials}

The five-pointed star orbit in the $r^5$
potential is in some ways a natural ``sweet spot'' for scarring. The $\reson{2}{5}$ resonance produces
tight near-degenerate resonant sets, because moving from one resonant state to
the next involves the exchange of a small number of quanta, making the
degeneracy approximation better.

The five-fold symmetry also gives more room for visually distinct preferred
orientations, as opposed to, e.g., an 11-pointed star. Moreover, the existence
of a self-crossing point in the classical orbit makes the probability density
in a scar less uniform, which helps the localization of the scars as they will
pin more strongly to impurities near the self-crossing point. This gives the
five-pointed star some advantage over the two simpler resonances, $\reson{1}{3}$ and
$\reson{1}{4}$.

In other homogeneous potentials, scarring due to classical resonances exists
analogously to the $r^5$ case. With other choices of the exponent in the
unperturbed potential, different resonances exist, as summarized in
Table~\ref{tab:raorbits}. This is reflected in the shape and abundance of
scars. Figure~\ref{fig:penta-examples} includes examples of scars in a $r^8$
potential.

The integer exponent $a=3$ is unusual in that its simplest non-trivial resonance
is $\reson{5}{11}$. With coefficients so large the near-degeneracy in the resonant sets
becomes poor, and the scars become rare and distorted. In fact, at this limit
the near-degeneracy due to resonant sets might compete with purely
accidental near-degeneracies. An even more extreme absence of short non-trivial classical
resonances can be found with non-integer exponents close to~$2$.

\section{Scars in a non-homogeneous potential}

While a homogeneous potential function simplifies the classical PO structure,
it is not necessary for scarring. Other forms of the unperturbed
potential also have quasi-degenerate sets of eigenstates due to classical
resonances, and thus local impurities will cause scars.
Figure~\ref{fig:penta-examples} contains an example of a
five-pointed scar eigenstate for $V(r) \propto \cosh(r)-1$. In this case, the
parameters of the perturbation are such that each Gaussian bump has
amplitude~$M=10$, FWHM of~$0.353$, and the average density of bumps is~$2.4$
bumps per unit square.

To provide more proof that similar scarring also exists for a non-homogeneous
potential, Fig.~\ref{fig:cosh-angle-comparison} shows wave-packet recurrences
as a function of the wave-packet orientation angle, and
Fig.~\ref{fig:cosh-peaks} shows eigenstates contributing to the recurrences of
the strongest preferred orientation.

\begin{figure}[htbp]
	\centering
	\includegraphics[width=\linewidth]{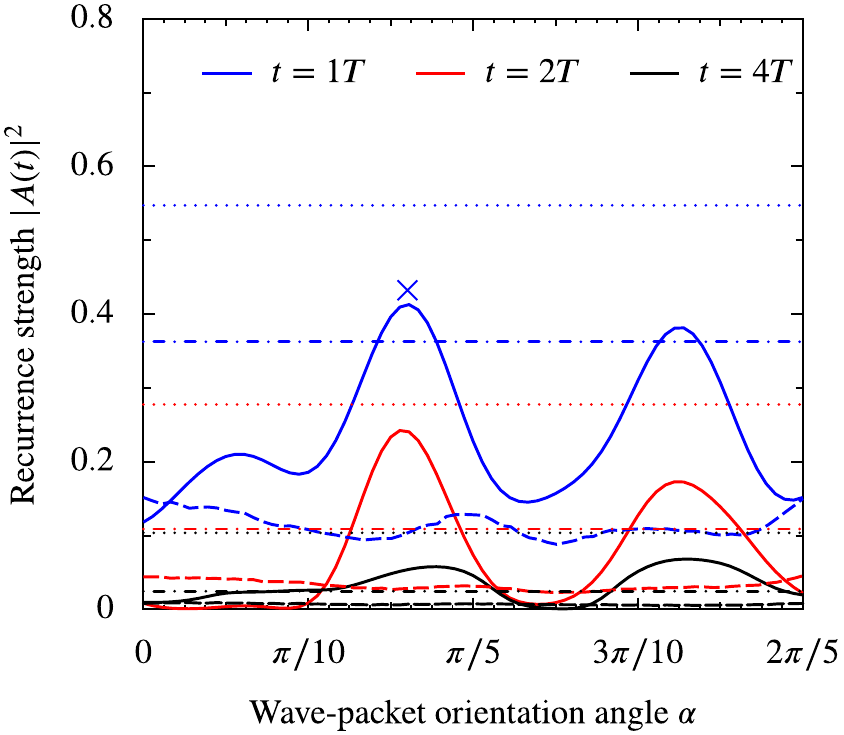}
	\caption{Plot equivalent to Fig.~\ref{fig:angle-comparison},
	except for a non-homogeneous unperturbed potential $V(r) \propto
	\cosh(r)-1$, with impurity parameters described in the text. The
	wave-packet was initialized on a $\reson{2}{5}$ PO with
	energy~$E=200$.}
	\label{fig:cosh-angle-comparison}
\end{figure}

\begin{figure}[htbp]
  \centering
  \includegraphics[width=0.90\linewidth]{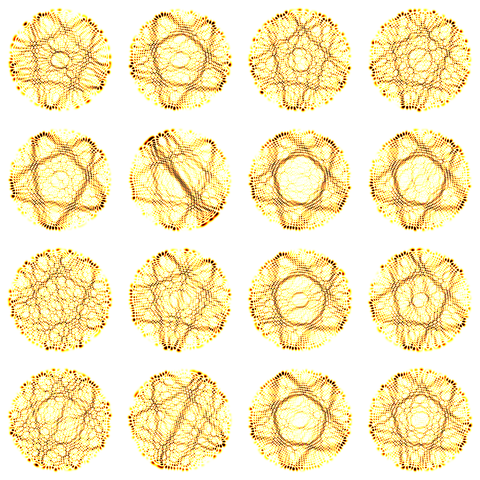}
  \caption{Most prominent eigenstates in the decomposition of the wave-packet
  in Fig.~\ref{fig:cosh-angle-comparison} at the most strongly recurring
  orientation (marked with a blue cross in Fig.~\ref{fig:cosh-angle-comparison}).}
  \label{fig:cosh-peaks}
\end{figure}

\section{Generality of the results for different random realizations of the impurities}
\label{supp:otherrealizations}

We have not attempted to specify quantitatively how common the scars or
preferred orientations are among all random realizations of the impurity
locations. To illustrate that they are not a rare occurrence by any means,
Fig.~\ref{fig:A24-copies} shows plots equivalent to
Fig.~\ref{fig:fair-overlaps}, but with 10 randomly selected
realizations of the impurity potential, each with the same parameters. The
realizations were generated by seeding the random number generator (RNG) in
\texttt{itp2d}~\cite{itp2d} with integers from~1 to~10.

While the preferred
direction branches are not always as clear as in the example realization
used in the main article, preferred orientations and accompanied strongly
scarred states (signified by the large overlap with the wave packet) are
present in all the shown cases.

\begin{figure*}[htbp]
  \centering
  \includegraphics[width=\textwidth]{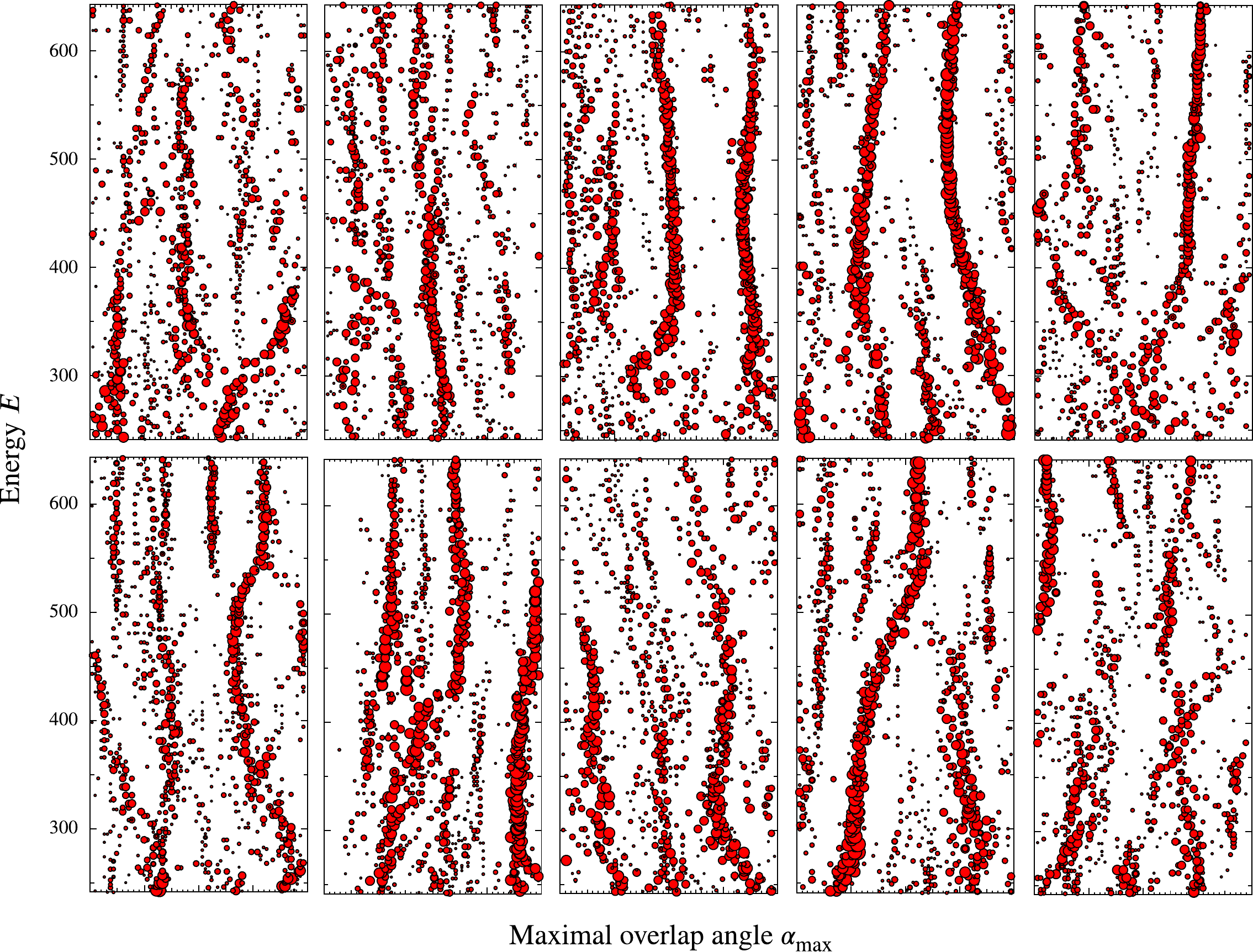}
  \caption{Montage of plots equivalent to Fig.~\ref{fig:fair-overlaps}, except
  than with ten other random realizations of the impurity positions. Each case
  shows branches of preferred orientations consisting
  of scarred eigenstates, although the strength of the branches and their
  stability with respect to eigenstate energy varies. In all plots the range of
  $\alpha_\text{max}$ is from $0$ to $2\pi/5$.}
  \label{fig:A24-copies}
\end{figure*}

The RNG seed used for the realization discussed in the main article is 20141010.
In the interest of full
reproducibility, the version of \texttt{itp2d} used was
\texttt{1.0.0-7-gd3c0454}, with command line parameters:

{\ttfamily%
\noindent itp2d --rngseed 20141010\\
\hspace*{1em}-F abschange(1e-3) -T absstdev(1e-3,5e-4)\\
\hspace*{1em}-l 11 -s 300 -e 0.01 -d 12 -t 6\\
\hspace*{1em}--noise impurities\\
\hspace*{1em}--impurity-distribution "uniform(2.0)"\\
\hspace*{1em}--impurity-type "gaussian(24, 0.1)"\\
\hspace*{1em}-n 4000 -N 5000\\
\hspace*{1em}-p "poweroscillator(5)"\\
\hspace*{1em}--maxsteps 50 --recover -D 2%
}

}{}

\end{document}